\documentclass[onecollarge]{svjour2}
\smartqed  % flush right qed marks, e.g. at end of proof
\usepackage{graphicx}
\usepackage{mathptmx}
\usepackage{amssymb}
\usepackage{amsmath}
\usepackage{graphicx}% Include figure files
\usepackage{bm}
\journalname{Few-Body Systems}
\begin{document}

\title{Weakly bound Li He$_2$  molecules\thanks {This work
was supported in part by the Heisenberg-Landau Program and the
Russian Foundation for Basic Research.}}

\titlerunning{Weakly bound Li He$_2$  and He$_3$ molecules}     % if too long for running head

\author{E. A. Kolganova  }

\authorrunning{E.\,A.\,Kolganova} % if too long for running head

\institute{E. A. Kolganova   \at
              Bogoliubov Laboratory of Theoretical Physics, Joint Institute for Nuclear Research and  Dubna State University, 141980 Dubna, Russia
          }

\date{Received: date / Accepted: date}

\maketitle

\begin{abstract}
The results for binding energies of $^6$Li He$_2$  and $^7$Li He$_2$ systems are presented. They are obtained by solving Faddeev equations in
configuration space. It is shown that the excited  states in both systems are of the Efimov-type.
\end{abstract}

\bigskip

\section{Introduction}

In the last few years an enormous progress was demonstrated in the studying of the Efimov effect~\cite{VEfimovYaF1970} in different systems. First of all it concerns ultracold quantum gases trapped by a magnetic field. Being a subject to a magnetic field, certain two-atom systems experience a Feshbach resonance due to Zeeman interaction. In such a case one gets an opportunity to control the atom-atom scattering length, by changing the intensity of the magnetic field.
The Efimov effect occurs if the two-body scattering length is large compared to the radius of the two-body interaction, then the three-body system may have an infinite number of weakly bound states. The energies of the Efimov levels are universally related and this relation does not depend on the form of the pair-wise interactions in the three-body system~\cite{VEfimovYaF1970}.
 
In 2006 there was the first observation of Efimov-type resonance in an ultracold gas of cesium atoms~\cite{Kraemer}. The resonance has to occur in the range of large negative two-body scattering lengths, arising from the coupling of three free atoms to an Efimov trimer. Experimentally, in \cite{Kraemer}  its signature was observed as a giant three-body recombination loss when the intensity of the magnetic field and so the strength of the two-body interaction was varied. Striking manifestations of the Efimov effect have been predicted for three-body recombination processes in ultracold gases with tunable two-body interactions in \cite{Esry99,Braaten}.
Although in this experiment only one Efimov resonance was observed, recently the second Efimov level has been measured using the same technique\cite{HuangGrimm2014}.
Starting from the first experiment by R. Grimm's group a lot of other experimental evidence for the Efimov states in three-atomic systems consisting of Li, K, Rb, Cs atoms and its combinations were reported \cite{Barontini2009,Lompe2010,Williams6Li,Rem7Li_2013,Bloom40K87Rb,Ulmanis6LiCs}.

Universality of Efimov effect allows to expect its manifestation  in nuclear systems, especially in a halo nuclei. Although the experimental evidence is not yet found, the existence of nuclear Efimov trimer states has been speculated in many isotopes \cite{Riisager2013}. Of particular interest are the investigations of the isotop of $^{22}$C, recently studied in~\cite{Ershov,Acharya2013,Gridnev2016}, which has so far the largest detected halo formed by two neutrons~\cite{Tanaka2010}, and of the isotop $^{62}$Ca -- the heaviest Borromean nucleus~\cite{Hagen2013}. Studying these systems improves our understanding of nuclear behavior in extreme conditions along the neutron drip line \cite{Kemper2010}.

One of the best theoretically predicted examples of the Efimov three-body system is a naturally existing molecule  of the helium trimer where an excited state is of the Efimov nature (see,~\cite{FBS2011} and refs. therein). Only recently there was the first observation of this long-predicted %Efimov 
state of helium trimer using the combination of  Coulomb explosion imaging with cluster mass selection by matter wave diffraction~\cite{Kunitski}.
%Most recently this has brilliantly been confirmed experimentally~\cite{Kunitski}.
The interaction between two helium atoms is quite small and supports only one bound state with the energy about 1mK and so a very large scattering length around 100 \AA. In addition to the Helium dimer, the He - alkali-atom interactions are even shallower and also support weakly bound states.
Thus, in triatomic $^4$He$_2$ - alkali-atom system one can expect the existence of Efimov levels. The  $^{6;7}$Li$^4$He$_2$ systems, which are investigated in this work, have the excited states of Efimov-type as will be demonstrated further by the results of the calculations. 

%The paper is organized as follows.
%In Sect. \ref{Sdt} we describe the Faddeev equations used for
%calculation of tritomic asymmentric system. In Sect. \ref{Salk} 
%we briefly discuss the obtained
%works on Efimov physics in ultracold alkali-atom gases.

\section{Method}%, Results and Discussions}
\label{Sdt}

We solve the Faddeev equations for three interacting atoms using a similar scheme as in our previous investigations of Helium trimer \cite{KMS-JPB}. There, the formalism which consists of a hard-version of the Faddeev differentional equations has been described in detail. Since the method employed is mostly the same as detailed in Refs. \cite{KMS-JPB} and \cite{FBS2004}  we give only a brief outline here.

In the present investigation we assume that LiHe$_2$ molecule has a total angular momentum $L=0$. 
Describing the $^7$Li\,$^4$He$_2$ three-atomic system we use the standard reduced Jacobi coordinates~\cite{MF}.
We consider the case where the interatomic interactions include a hard core component. Outside the hard core
domain they are described by conventional smooth potentials. 
In the following the $^4$He atoms are assigned the numbers 1 and 2 while the  $^7$Li atom has the number 3. The identity of the two $^4$He atoms implies that the corresponding Faddeev components are obtained from each other by a simple rotation of the coordinate space.
Thus, we only have two independent Faddeev components, the one associated with the $^4$He--$^4$He subsystem,
and another one associated with a pair of $^7$Li and $^4$He atoms.
After partial-wave expansion the initial Faddeev equations~\cite{MF} are reduced to the system of coupled two-dimensional 
integro-differential equations
\begin{eqnarray}
\nonumber
&{\left(-\displaystyle\frac{\partial^2}{{\partial x}^2}-
\frac{\partial^2}{{\partial y}^2}+
l(l+1)\left(\displaystyle\frac{1}{x^2}+\frac{1}{y^2}\right)-E
\right)f^{(\alpha)}_l(x,y)}\\
\label{Fparts}
& \qquad=\left\{\begin{array}{cc} 0, & x<c \\
-V_\alpha(x)\psi_l^{(\alpha)}(x,y), & x>c
\end{array}\right., \qquad \alpha=1,3,
\end{eqnarray}
and partial boundary conditions
\begin{equation}
\label{HCparts}
 \left.f^{(\alpha)}_l(x,y)\right|_{x=0}= \left.f^{(\alpha)}_l(x,y)\right|_{y=0}=0, \quad \left.\psi_l^{(\alpha)}(x,y)\right|_{x=c}=0,\quad \alpha=1,3.
\end{equation}
Here, $x$ and $y$ stands for reduced Jacobi variables. $\psi_l^{(\alpha)}(x,y)$ are the partial wave functions related to the 
partial-wave Faddeev amplitudes $f^{(\alpha)}_l(x,y)$ (see, e.g.~\cite{FBS2004}):
\begin{eqnarray*}
\psi_l^{(\alpha)}(x,y)&=&f_l^{(\alpha)}(x,y)+\displaystyle\sum\limits_{l',\beta \ne \alpha}\int\limits_0^1 d\eta
h^0_{(\alpha;l)(\beta;l')}(x,y,\eta)f^{(\beta)}_{l'}(x_{\beta \alpha}(\eta),y_{\beta \alpha}(\eta)).
\end{eqnarray*}
The explicit form of the function $h^0_{(\alpha;l)(\beta;l')}(x,y,\eta)$ can be found in Refs.\cite{FBS2004,MF}.
We also use the notation
\begin{eqnarray*}
x_{\beta\alpha}(\eta)&=&\sqrt{{\sf c}_{\beta\alpha}^2 x^2+
2{\sf c}_{\beta\alpha}{\sf s}_{\beta\alpha} xy\eta
+{\sf s}_{\beta\alpha}^2 y}, \quad
y_{\beta\alpha}(\eta)=\sqrt{{\sf s}_{\beta\alpha}^2 x
-2{\sf c}_{\beta\alpha}{\sf s}_{\beta\alpha} xy\eta
+{\sf c}_{\beta\alpha}^2 y}.\\
 {\sf c}_{\alpha\beta}&=&-\left(
\frac{{\rm m}_\alpha {\rm m}_\beta}
    {({\rm m}_\alpha + {\rm m}_\beta)
    ({\rm m}_\beta + {\rm m}_\gamma)} \right)^{1/2}, \quad
    {\sf s}_{\alpha\beta}=(-1)^{\beta-\alpha}
\mathop{\rm sign}(\beta-\alpha)
\left( 1-{\sf c}_{\alpha\beta}^2 \right)^{1/2}.
\end{eqnarray*}
where  ${\sf c}_{\alpha\beta}$ and ${\sf s}_{\alpha\beta}$ stand for the angular coefficients describing the transition from
the reduced Jacobi variables associated with a pair $\beta$ to the ones associated with a pair $\alpha$.
By $c$ in equations (\ref{Fparts}), (\ref{HCparts}) we denote the hard-core radius. This radius was taken the same for all three inter-atomic interaction potentials and was chosen in such a way that any further decrease of it does not affect the trimer ground-state energy. 
A detailed description of the Faddeev differential equations in the hard-core
model in case of symmetric helium trimer can be found in \cite{KMS-JPB}.
By $V_1$  we denote the interatomic Li--He potential and $V_3$ - the He--He
potential adjusted to the corresponding reduced Jacobi coordinates.

The asymptotic boundary condition for a LiHe$_2$
bound state reads as follows (see \cite{FBS2004,MF})
\begin{equation}
\label{HeBS}
        \begin{array}{rcl}
  f^{(\alpha)}_l(x,y) & = & 
  \delta_{l0}\psi_d(x)\exp({\rm i}
  \sqrt{E-\epsilon_d}\,y)
    \left[{\rm a}_0+o\left(y^{-1/2}\right)\right] \\
        && + \displaystyle\frac{\exp({\rm i}\sqrt{E}\rho)}{\sqrt{\rho}}
      \left[A^{(\alpha)}_l(\theta)+o\left(\rho^{-1/2}\right)\right]
\end{array}
\end{equation}
as $\rho=\sqrt{x^2+y^2}\rightarrow\infty$ and/or
$y\rightarrow\infty$.  The coefficients ${\rm a}_0$ and $A^{(\alpha)}_l(\theta)$ describe contributions into
$f^{(\alpha)}_l(x,y)$ from (2+1) and (1+1+1) channels respectively. It should be noted that both $E-\epsilon_d$
and $ E$ in (\ref{HeBS}) are negative. This implies that for any $\theta=\arctan(y/x)$ the  
partial wave Faddeev amplitudes $f^{(\alpha)}_l(x,y)$ decrease exponentially as $\rho\rightarrow\infty$ .
Here we also use the fact that dimers, $^4$He$_2$ and Li$^4$He, have a unique bound state and this state only
exists for $l=0$;  $\epsilon_d$ stands for the correspondent 
dimer energy while $\psi_d(x)$ denotes the dimer
wave function which is assumed to be zero within the core,
that is, $\psi_d(x)\equiv 0$ for $x\leq c$.

Here we only deal with a finite number of equations
(\ref{Fparts}), assuming $l \leq l_{\rm max}$ where $l_{\rm max}$ is a certain non-negative
integer. As in \cite{KMS-JPB,FBS2004} we use a
finite-difference approximation of the boundary-value
problem (\ref{Fparts}), (\ref{HCparts}), (\ref{HeBS}) in the polar
coordinates $\rho$ and $\theta$. The grids are chosen
so that the points of intersection of the arcs
$\rho=\rho_i$, $i=1,2,\ldots, N_\rho$ and the rays
$\theta=\theta_j$, $j=1,2,\ldots, N_\theta$ with the
core boundary $x=c$ constitute the  nodes. The value of
the core radius is chosen to be $c=1$\,{\AA} by the
argument given in \cite{MSSK}. We also follow the same
method for choosing the grid radii $\rho_i$ (and, thus,
the grid hyperangles $\theta_j$) as described in
\cite{KMS-JPB,MSSK} in details. Atomic mass of isotops are taken from~\cite{Mills}.

%\begin{wrapfigure}[19]{r}{0.5\textwidth}
%	\centering \hspace*{-0.1truecm} {\small $V$ (K)}
%	    \includegraphics[width=0.43\textwidth]{Potentials.png}
%		\hspace*{5.truecm} {\small $r$ ({\AA})}
 % \caption{The He--He LM2M2 potential and  He - alkali-atoms KTTY potentials $V$ (in K)
%as a function of the interatomic distance $r$ (in \AA)}
  %\label{Fig-Pot}
%\end{wrapfigure}

\section{Results and Discussions}
\label{Salk}

Our calculations are based on the semi-empirical LM2M2 potential~\cite{Aziz91} proposed by
Aziz and Slaman for He-He interaction, and the  KTTY potential~\cite{KTTY} , theoretically derived  by  Kleinekath\"ofer, Tang,
Toennies and Yiu  for Li-He interaction with more accurate coefficients taken from \cite{Yan,KLM-KTTY}. 
Both of these potentials are widely used in the literature. Calculated values of the binding energy for
$^6$Li$^4$He is 1.512 mK  and for $^7$Li$^4$He is 5.622 mK.  Such small values of binding energy give indication on possible existence of Efimov states in corresponding He$_2$ - alkali-atom triatomic systems.

 We employed the equations (\ref{Fparts}),
(\ref{HCparts}) and the bound-state asymptotic boundary
condition (\ref{HeBS}) to calculate the binding energy of
the trimer Li$^4$He$_2$. The three-body interaction is expected to be small as in the case of helium trimer~\cite{Cencek}
and we do not take it into account. Our results for the $^{6;7}$Li$^4$He$_2$
trimers binding energies, as well as the results obtained by other authors, are presented in Table \ref{tableBS}.
The results show that used potential models support two bound states. The energy of excited state is 
very close to the LiHe two-body threshold. However, as it is seen from the Table~\ref{tableBS}, different methods demonstrate a large discrepancy between the results. 
In contrast to our calculation, the hyperradial Shr\"odinger equation %in hyperspherical coordinates 
has been solved by other authors.
The third column contains the results obtained by Wu {\it et al.} \cite{Wu} using the mapping method within the frame of the hyperspherical coordinates~\cite{Kokoouline}. The next two columns are the results of calculations by H. Suno,  E. Hiyama and M. Kamimura~\cite{Suno13} using 
the Gaussian expansion method and the adiabatic hyperspherical representation respectively, although with different He-He potentals. They employed
the He-He potential suggested by Jeziorska \textit{et al.}~\cite{Jeziorska}. The two methods are found to differ from each other, but authors in \cite{Suno13}
mentioned that the adiabatic hyperspherical representation is less accurate.
The next column is the results of calculations by Suno and Esry~\cite{Suno09,Suno10} by the adiabatic hyperspherical method. They also employed
the He-He potential from~\cite{Jeziorska}, but different potential for Li-He interaction proposed by Cvetko  \textit{et al.}~\cite{Cvetko}. The seventh column contains the results obtained by Baccarelli  \textit{et al.}~\cite{Baccarelli} with the same potential as in~\cite{Suno10}, but using a different computation method.  The column VIII presents one of the first results obtained by Yuan and Lin~\cite{Yuan}, using the adiabatic hyperspherical method which gives an upper bound to the ground state. In the last column is the prediction of the bound state energies made by Delfino \textit{et al.}~\cite{Delfino} using the scaling ideas and zero-range model calculations.
We can see from the Table~\ref{tableBS} that all calcultations predict the existence of two states in both $^6$Li$^4$He$_2$ and $^7$Li$^4$He$_2$systems. The energy of the excited state is close to the two-body LiHe threshold which is lower then He$_2$. 
The binding energies are very sensitive to the methods and the potential models used as it was aslo mentioned in~\cite{Wu,Suno10}.
However, the bound-state energy for the nonsymmetrical helium trimer $^3$He$^4$He$_2$ obtained using adiabatic hyperspherical approach~\cite{Suno08} agrees fairly well with our previous calculations of this system~\cite{FBS2004}. For the potentials used, the difference in these systems is in the larger value of the binding energy of the $^7$LiHe dimer than the He$_2$ dimer. It means that the system $^7$LiHe$_2$ is not so close to the universal regime as in case of  the helium trimer and it could be the reason of the discrepancy.

%%%%%%%%%%%%%%   TABLE III: Sc.Len.  %%%%%%%%%%%%%%%%%%%%%%%%%%%%%%
\begin{table}[t]
%\beforetab
\caption{Comparison of the bound state energies (in mK) obtained  for a grid with $N_\rho =N_\theta= 800$, $\rho_{max}$ up to $1000$ \AA\  and $l_{max}=4$  with other calculations.}
\label{tableBS}
\begin{center}
\begin{tabular}{|c|c|c|c|c|c|c|c|c|}
\hline
%\postline
% &  &  \multicolumn{7}{c}{} \\
%\preline
%\cline{3-9}%\postline
 E (mK)      & present &~\cite{Wu} &~\cite{Suno13} & ~\cite{Suno13} & ~\cite{Suno10} & ~\cite{Baccarelli} &~\cite{Yuan} &~\cite{Delfino} \\
He-He potential& LM2M2 & LM2M2 & Jeziorska & Jeziorska & LM2M2 & LM2M2 & KTTY & \\
He-Li potental & KTTY & KTTY & KTTY & KTTY &Cvetko & Cvetko & KTTY & \\
% & & & & & & & &\\
\hline
$|$E$_{^{7}Li^{4}He_{2}}|$ & 50.89 & 78.73 & 76.32 &81.29 & 64.26  & 73.3 & 45.7 & 45.7   \\
$|$E$^{*}_{^{7}Li^{4}He_{2}}|$  & 5.625 & 5.685 & 5.51 & 5.67  & 3.01 & 12.2 &   & 2.31 \\[-2ex]
%$|$E$^{*}_{^{7}Li^{4}He_{2}}$ -- $\epsilon_{d}$$|$  & 0.003 & 0.063 & 0.05 & 0.47 & 9.42 &   & 0.15 \\
 & & & & & & & &\\
\hline
$|$E$_{^{6}Li^{4}He_{2}}|$ & 35.45 &  &  &58.88&   & 51.9 & 31.4 & 31.4   \\
$|$E$^{*}_{^{6}Li^{4}He_{2}}|$  & 1.719 &  &  &2.09 &  & 7.9 &    & - \\[-2ex]
 & & & & & & & &\\
\hline
\end{tabular}
\end{center}
\end{table}

The excited state of $^7$LiHe$_2$  demonstrates a Efimov-type behavior.  To study the Efimov properties we multiplied the original Li-He potential by a factor $\lambda$. An increase of the coupling constant $\lambda$ makes potential more attractive. In this case the Efimov levels should become weaker and disappear with further increase of $\lambda$. Namely this situation is observed for the excited state energy of $^7$LiHe$_2$ in contrast to the ground state energy whose absolute value increases continuously  with increasing attraction. The difference between the dimer energy of $^7$LiHe (the lowest two-body threshold) and the energy of the $^7$LiHe$_2$ trimer excited state increases with potential weakening up to the moment when the energy of the $^7$LiHe dimer become less than the energy of He$_2$. Further decrease of the coupling constant weakens only the Li-He potential and although the LiHe dimer energy is approaching zero, the He-He two-body threshold remain the same. So the difference between the He$_2$ dimer energy  and the energy of the $^7$LiHe$_2$ trimer excited state becomes smaller and the excited state disappears with further decreasing of $\lambda$. As it was shown for helium trimer, the Efimov level transforms into a virtual state~\cite{He3-Rev} .  It would be interesting to see what happens in the case of the LiHe$_2$ system and it is a subject of our further investigations.

\begin{acknowledgements}
The author would like to thank P. Stipanovi\'c and A.Kievsky for stimulating discussions and also W.~Sandhas and A.K.~Motovilov 
for their constant interest to this work.
\end{acknowledgements}


\begin{thebibliography}{00}

\bibitem{VEfimovYaF1970}
V. N. Efimov,  {\it Weakly-bound states of 3 resonantly-interacting particles}, Sov. J. Nucl.
Phys. \textbf{12} (1970), 589 [Yad. Fiz. \textbf{12} (1970), 1080].

%\bibitem{VEfimov1970}
%V. Efimov, Phys. Lett. B \textbf{33} (1970), 563.

%\bibitem{VEfimov1973}
%V. Efimov,  Nucl. Phys. A. \textbf{210} (1973), 157.


\bibitem{Kraemer}
T. Kraemer, M. Mark, P. Waldburger, J. G. Danzl, C. Chin, B.
Engeser, A. D. Lange, K. Pilch, A. Jaakkola, H.-C. Na\"gerl, and R.
Grimm, 
{\it Evidence for Efimov quantum states in an ultracold
gas of caesium atoms},
Nature \textbf{440} (2006), 315.

\bibitem{Esry99}
B. D. Esry,   C. H. Greene,  and J. P. Burke, 
{\it Recombination of Three Atoms in the Ultracold Limit}, Phys. Rev. Lett.
\textbf{83} (1999), 1751.

\bibitem{Braaten}
E. Braaten and H.-W. Hammer, 
{\it Universality in few-body systems with large scattering length},
Phys. Rep. \textbf{428} (2006), 259.

\bibitem{HuangGrimm2014}
B.  Huang,   L. A. Sidorenkov,  and R. Grimm, 
{\it Observation of the Second Triatomic Resonance in Efimov’s Scenario},
Phys. Rev. Lett.
\textbf{112} (2014), 190401.


\bibitem{Barontini2009}
G. Barontini, C. Weber, F. Rabatti, J. Catani, G. Thalhammer, M.
Inguscio, and F. Minardi, 
{Observation of Heteronuclear Atomic Efimov Resonances},
Phys. Rev. Lett. \textbf{103} (2009),
043201; Erratum, Ibid. \textbf{104} (2010), 059901.

\bibitem{Lompe2010}
T. Lompe, T. B. Ottenstein, F. Serwane, K. Viering, A. N. Wenz, G.
Z\"urn, and S. Jochim, 
{\it Radio-Frequency Association of Efimov Trimers}
Science \textbf{330} (2010), 940.
%Phys. Rev. Lett. \textbf{105} (2010), 103201.

\bibitem{Williams6Li}
J. R. Williams, E. L. Hazlett, J. H. Huckans, R. W. Stites, Y.
Zhang, and K. M. O'hara, 
{\it Evidence for an Excited-State Efimov Trimer in a Three-Component Fermi Gas},
Phys. Rev. Lett.  {\bf 103} (2009), 130404.


 \bibitem{Rem7Li_2013}
B. S. Rem, A. T. Grier, I. Ferrier-Barbut, U. Eismann, T. Langen, N.
Navon, L. Khaykovich, F. Werner, D. S. Petrov, F. Chevy, and C.
Salomon, 
{\it Lifetime of the Bose Gas with Resonant Interactions},
Phys. Rev. Lett. {\bf 110} (2013), 163202.


\bibitem{Bloom40K87Rb}
R. S. Bloom, M.-G. Hu, T. D. Cumby, and D. S. Jin 
{\it Tests of Universal Three-Body Physics in an Ultracold Bose-Fermi Mixture},
Phys. Rev. Lett.
{\bf 111}  (2013), 105301.

\bibitem{Ulmanis6LiCs}
J. Ulmanis, S. H\"afner, R. Pires, F. Werner, D. S. Petrov, E. D.
Kuhnle, and M. Weidem\"uller, 
{\it Universal three-body recombination and Efimov resonances in an ultracold Li-Cs mixture},
Phys. Rev. A {\bf 93} (2016), 022707.

\bibitem{Riisager2013}
K. Riisager,  
{\it Halos and related structures},
Physica Scripta \textbf{T152} (2013), 014001.

\bibitem{Ershov}
S. N. Ershov, J. S. Vaagen, and M. V. Zhukov,
{\it Binding energy constraint on matter radius and soft dipole excitations of $^22$C},
Phys. Rev. C \textbf{86} (2012), 034331.

\bibitem{Acharya2013}
B. Acharya, C. Ji, D. Phillips,  
{\it Implications of a matter-radius measurement for the structure of Carbon-22},
Phys. Lett. \textbf{723} (2013), 196.

\bibitem{Gridnev2016}
D. Gridnev, D. Bressanini, 
{\it Manifestation of Universality in the Asymmetric
Helium Trimer and in the Halo Nucleus $^{22}$C},
arXiv:1604.05854

\bibitem{Tanaka2010}
K. Tanaka, T. Yamaguchi, T. Suzuki, {\em \ et al.} 
%T. Ohtsubo, M. Fukuda, D. Nishimura, M. Takechi, K. Ogata, A. Ozawa,
%T. Izumikawa, T. Aiba, N. Aoi, H. Baba, Y. Hashizume, K. Inafuku, N. Iwasa, K. Kobayashi, M. Komuro,
%Y. Kondo, T. Kubo, M. Kurokawa, T. Matsuyama, S. Michimasa, T. Motobayashi, T. Nakabayashi, S. Nakajima,
%T. Nakamura, H. Sakurai, R. Shinoda, M. Shinohara, H. Suzuki, E. Takeshita, S. Takeuchi, Y. Togano,
%K. Yamada, T. Yasuno, and M. Yoshitake
{\it Observation of a Large Reaction Cross Section in the Drip-Line Nucleus $^22$C},
Phys. Rev. Lett.
\textbf{104} (2010), 062701.

\bibitem{Hagen2013}
G. Hagen, P. Hagen, H.W. Hammer, and L. Platter, 
{\it Efimov Physics Around the Neutron-Rich $^60$Ca Isotope},
Phys. Rev. Lett.
\textbf{111} (2013), 132501.

\bibitem{Kemper2010}
K. W. Kemper, P. D. Cottle, 
{\it A breakthrough observation for neutron dripline physics},
Physics {\bf 3} (2010), 13.

\bibitem{FBS2011}
E.\,A.\,Kolganova,  A.\,K.\,Motovilov, and W.\,Sandhas, 
{\it The $^4$He Trimer as an Efimov System},
Few-Body Syst. \textbf{51} (2011), 249.

\bibitem{Kunitski}
{\rm M. Kunitski}, S. Zeller, J. Voigtsberger,
A. Kalinin,L.Ph.H.Schmidt, M. Sch\"offer, A.Czasch,W. Sch\"ollkopf,
R. E. Grisenti,C. Janke, D.Blume, and R. D\"orner,
{\it Observation of the Efimov state of the helium trimer},
{\rm Science \bf 348} (2015),  551.

\bibitem{KMS-JPB}
E. A. Kolganova, A. K. Motovilov, and S. A. Sofianos, 
{\it Three-body configuration space calculations with hard-core
potentials},
J. Phys. B {\bf 31} (1998), 1279.

				
\bibitem{FBS2004}
W. Sandhas, E. A. Kolganova, Y. K. Ho, and A.K. Motovilov,
{\it Binding Energies and Scattering Observables in the $^4$He$_3$ and $^3$He$^4$He$_2$ Three-Atomic Systems},
Few-Body Syst. {\bf 34} (2004), 137.

\bibitem{MF}
       L. D.  Faddeev and S. P. Merkuriev,
       {\it Quantum scattering theory for several particle systems}
       (Kluwer Academic Publishers, Doderecht, 1993).		

\bibitem{MSSK} A. K. Motovilov, W. Sandhas, S. A. Sofianos, and
E. A. Kolganova, 
{Binding energies and scattering observables in the $^4$He$_3$ atomic system},
Eur. Phys. J. {\bf D 13} (2001), 33.

\bibitem{Mills}
I. Mills, T. Cvita\^s, K.Homann, N. Kallay, K. Kuchitsu, {\it Quantities, Units and Symbols in Physical Chemistry}, 2nd edition, Blackwell Science, Oxford, 1993.

\bibitem{Aziz91}
R. A. Aziz and  M. J. Slaman, 
{\it An examination of ab inifio results for the helium potential energy curve},
J. Chem. Phys. \textbf{94} (1991), 8047.
			
					
\bibitem{KTTY} U. Kleinekath\"ofer, K.T. Tang, J.P. Toennies, C.I.Yiu, 
{\it Potentials for some rare gas and alkali-helium systems
calculated from the surface integral method},
Chem. Phys. Lett. {\bf 249} (1996), 257. 

\bibitem{Yan}
Z.C. Yan, J. F. Babb, A. Dalgarno, and G.W.F. Drake, 
{\it Variational calculations of dispersion coefficients for interactions among H, He, and Li atoms},
Phys. Rev. A {\bf 54} (1996), 2824.

\bibitem{KLM-KTTY}
 U. Kleinekath\"ofer, M. Lewerenz, M.Mladenoc, 
{\it Long Range Binding in Alkali-Helium Pairs},
Phys. Rev. Lett. {\bf 83} (1999), 4717.


\bibitem{Cencek}
W. Cencek, M. Jeziorska, O. Akin-Ojo, and K. Szalewicz, 
{\it Three-Body Contribution to the Helium Interaction Potential},
J. Phys. Chem. A {\bf 111} (2007) (2007), 11311.

\bibitem{Wu}
M.-S. Wu, H.-L. Han, Ch.-B. Li, and T.-Y. Shi, 
{\it Prediction of a weakly bound excited state of Efimov character in a $^7$Li$^4$He$_2$ system},
Phys. Rev. A {\bf 90} (2014), 062506.

\bibitem{Kokoouline}
V. Kokoouline and F. Masnou-Seeuws, 
{\it Calculation of loosely bound levels for three-body quantum systems using hyperspherical
coordinates with a mapping procedure},
Phys. Rev. A {\bf 73} (2006), 012702.

\bibitem{Suno13}
H. Suno,  E. Hiyama, and M. Kamimura, 
{\it Theoretical Study of Triatomic Systems Involving Helium
Atoms},
Few-Body Syst. {\bf  54} (2013), 1557.

\bibitem{Suno09}
H. Suno and B.D. Esry, 
{\it Three-body recombination in cold helium–helium–alkali-metal-atom collisions},
Phys. Rev. A {\bf 80} (2009), 062702.

\bibitem{Suno10}
H. Suno and B.D. Esry, 
{\it Adiabatic hyperspherical study of weakly bound helium–helium–alkali-metal triatomic systems},
Phys. Rev. A {\bf 82} (2010), 062521.

\bibitem{Jeziorska}
M. Jeziorska, W. Cencek, K. Patkowski, B. Jeziorski, and K.
Szalewicz, 
{\it  Pair potential for helium from symmetry-adapted perturbation theory calculations and
from supermolecular data},
J. Chem. Phys. {\bf 127}  (2007), 124303.


\bibitem{Baccarelli}
I. Baccarelli, G. Delgado-Barrio, F.A. Gianturco, T. Conzalez-Lezana, S. Miret-Artes, and P. Villarreal, 
{\it Searching for Efimov states in triatomic systems: The case of LiHe$_2$},
Europhys. Lett. {\bf 50} (2000), 567.

\bibitem{Yuan}
J.M. Yuan and C.D. Lin, 
{\it Weakly bound triatomic He$_2$Li and He$_2$Na molecules},
J. Phys. B {\bf 31} (1998), L637.

\bibitem {Delfino}
A. Delfino, T. Frederico, L. Tomio, 
{\it Prediction of a weakly bound excited state in the $^4$He$_2$–$^7$Li molecule}
J. Chem. Phys. {\bf 113} (2000), 7874.


\bibitem{Cvetko}
D. Cvetko, A Lansi, A. Morgante, F. Tommasini, P. Cortana, and M.G. Dondi, 
{\it A new model for atom–atom potentials},
J. Chem. Phys. {\bf 100} (1994), 2052.

\bibitem{Suno08}
H. Suno and B.D. Esry, 
{\it Adiabatic hyperspherical study of triatomic helium systems},
Phys. Rev. A {\bf 78} (2008), 062701.


\bibitem{He3-Rev}
E.A. Kolganova,  A.K. Motovilov, and W. Sandhas,
{\it Ultracold collisions in the system of three helium atoms},
Physics of Particles and Nuclei \textbf{40},  206 (2009).
%206–-235


\end{thebibliography}
\end{document}